\documentclass[twocolumn]{article} \usepackage{graphicx}
\usepackage{bioinformatics} \usepackage{fullpage}

\newcommand{\PM}{{\it PM}} \newcommand{\MM}{{\it MM}}
\newcommand{\logA}{{\log(A^j_i)}} \newcommand{\isplit}{{i_{s}}}
\newcommand{\vmax}{{v_{\it max}}}

\begin{document}


\title{A study of accuracy and precision in oligonucleotide arrays:
extracting more signal at large concentrations}

\author{Felix Naef$^{1}$, 
Nicholas D. Socci$^{2}$ and 
Marcelo Magnasco$^{1}$\\
$^{1}$Center for Studies in Physics and Biology,
$^{2}$Laboratory for Molecular Genetics,\\
Rockefeller University, 
1230 York Avenue, NY 10021
}


\date{\today}

\maketitle
\begin{abstract}
\noindent\textbf{Motivation:}
Despite the success and popularity of oligonucleotide arrays as a high-throughput technique for measuring mRNA expression levels, quantitative calibration studies have until now been limited.
The main reason is that suitable data was not available. However, calibration data recently produced by Affymetrix now permits detailed studies of the intensity dependent sensitivity.
Given a certain transcript concentration, it is of particular interest to know whether current analysis methods are capable of detecting differential expression ratios of 2 or higher . 

\noindent\textbf{Results:}
Using the calibration data, we demonstrate that while current techniques are capable of detecting changes in the low to mid concentration range, the situation is noticeably worse for high concentrations.
In this regime, expression changes as large as 4 fold are severely biased, and changes of 2 are often undetectable. Such effects are mainly the consequence of the sequence specific binding properties of probes, and not the result of optical saturation in the fluorescence measurements.
{\sf GeneChips} are manufactured such that each transcript is probed by a set of sequences with a wide affinity range. We show that this property can be used to design a method capable of reducing the high intensity bias. The idea behind our methods is to transfers the weight of a measurement to a subset of probes with optimal linear response at a given concentration, which can be achieved using local embedding techniques.

\noindent\textbf{Availability:} Program source code will be sent
electronically upon request.

\noindent\textbf{Contact:} felix@funes.rockefeller.edu;\\
soccin@rockefeller.edu; marcelo@zahir.rockefeller.edu
\end{abstract}

\subsection*{Introduction}

High-density oligonucleotide arrays manufactured by Affymetrix are
among the most sensitive and reliable microarray
technology~\cite{GC2,GC3} available. Based on a photolithographic
oligonucleotide deposition process, labeled and amplified mRNA
transcripts are probed by 22-40 (depending on chip models)
short DNA sequence each 25 bases
long. The probes are preferentially picked near the 3' end of the mRNA
sequence, because of the limited efficiencies of reverse transcription
enzymes. In addition, the probes come in two varieties: half are perfect
matches (\PM\/) identical to templates found in databases, and the other
half are single mismatches (\MM\/), carrying a single base substitution
in the middle (13th) base of the sequence. \MM\ probes were introduced
to serve as controls for non-specific hybridization, and most analysis methods postulate that the actual signal (the target's mRNA concentration) to be proportional to the difference of match versus mismatch (\PM-\MM\/).

\begin{figure}[ht]
\centering
\includegraphics[width=1\columnwidth]{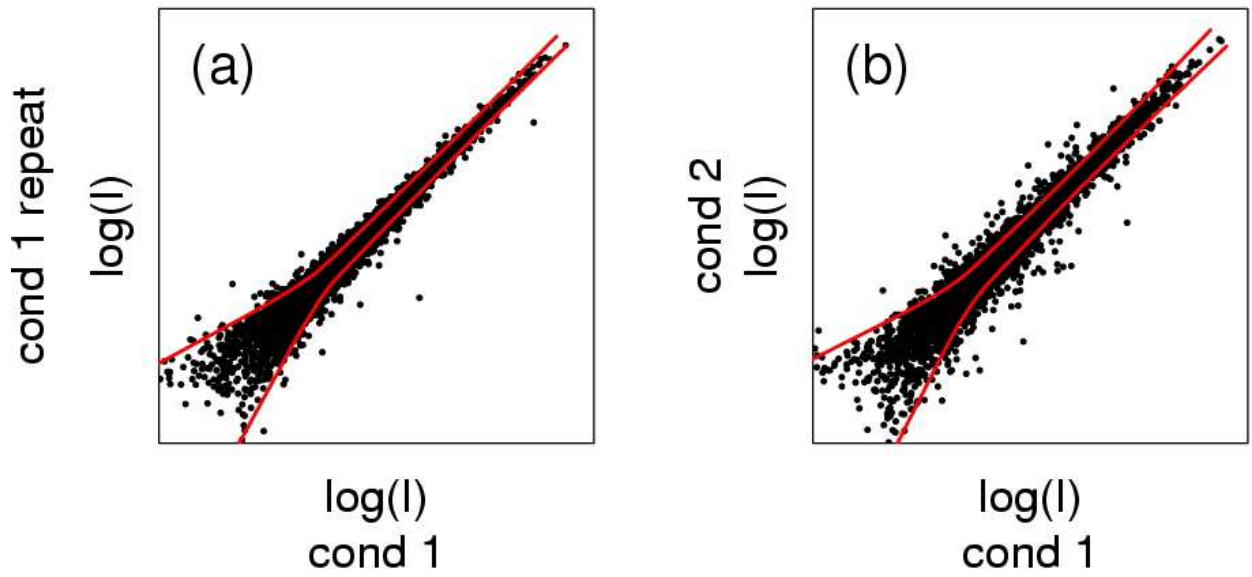}
\caption{\small Typical scatterplots from GeneChip data.  a) Log transformed
intensities for repeated hybridization conditions (duplicates).  b)
different conditions. The red lines show the lines of local standard deviation
(SD=2) in the log-rations.}
\label{fig:scatter}
\end{figure}

The purpose of this work is twofold. First, we present a detailed calibration study of {\sf GeneChips}. Specifically, we apply latest analysis methods (MAS 5.0 algorithm and others) to a large yeast calibration dataset, in which a number of transcripts are hybridized at known concentrations. We investigate the concentration dependence of both the accuracy and precision of differential expression scores.
By {\em accurate}, we mean that the reported numerical ratios values
are close to the known expression ratios, and have therefore little bias. On the other hand, a measurement is {\em precise} if it has a low 
noise level, also referred to as small variance.    
Our results show that the ability of conventional analysis techniques to detect small changes strongly deteriorates toward high transcript concentrations.
While the variance is smallest for high concentrations, it appears that the question of the bias in this regime has been neglected. In fact, the bias is strong enough that real changes of 2 (even 4) can often not be detected. 
This sounds at first counter-intuitive, which we believe is
rooted in the following widespread interpretation of hybridization data.
Namely, when examining the data from two replicated conditions
(Figure~\ref{fig:scatter}a)
%
%
%
%
most would focus on the low intensity region, and observe how noisy this regime appears to be in comparison to the high-intensity tail.
However, this view is misleading, as it does not consider the question of the bias.
Turning to a comparison of two different conditions (Figure~\ref{fig:scatter}b), we notice that the noise envelope is essentially unchanged, and that real changes appear as points lying distinctively outside the noise cloud.
Looking at multiple such comparisons, we would then conclude that the
high intensity data is almost always very tightly scattered about the
diagonal, and that there are rarely genes in that region that show
fold changes greater than, say, 1.5 or 2. The interpretation that no
differential regulation occurs in highly expressed transcripts seems
unlikely. In fact, we show evidence that real changes are often
compressed for large concentrations. This saturation effect can
actually be observed in Affymetrix's own data%
\footnote{Figure 7 at\\
{\tt http://www.affymetrix.com/products/algorithms\_tech.html}} 
although the issue is not commented there (a qualitative report has been given in~\cite{chudin}). 
The physical origin for the compression effect invokes non-linear
probe affinities and chemical saturation. This is a {\em separate}
issue from optical saturation (cf. Results).
Chemical saturation occurs below the detector threshold and is attributed to the fact that some probes will exhaust their binding capacities at relatively low concentrations, simply because their binding affinities are high.
Binding affinities are in fact very sensitive to the sequence
composition, resulting in measured brightnesses that usually vary by
several decades within a given probeset~\cite{US2}.

Our second goal is to present an analysis method that reduces the bias
at high concentrations. Our approach uses all \PM\ and \MM\/ probes equally, in contrast to the standard view in which the {\PM}s are thought to carry the signal, while the {\MM}s serve as non-specific controls. In fact, it has become clear that the \MM\ probes also track the signal, usually with lower (although often with higher) affinities than the \PM's \cite{US2}. In that sense, the \MM's should be viewed as a set of on average lower affinity probes. It is then reasonable to expect that some \MM\ probes will more accurate at high intensities, since they will be less affected by saturation than the the \PM's (cf. Figure ~\ref{fig:twoarray}).

\subsection*{Methods}

The existing methods for the analysis of the raw data fall into two
main classes. The first methods are similar to
Affymetrix's Microarray Suite software, providing absolute intensities
on a chip by chip basis, or differential expression ratios from 
two experiments~\cite{AFFYSOFT,US1,US3}. The second class are called
``model-based'' approaches~\cite{LIWONG}, and attempt to fit the probe
affinities from a large number of experiments.

The method described below belongs to the second class and is specifically designed for improved accuracy in the compressive high-intensity regime.
It is based on ideas borrowed from the theory of locally linear embeddings~\cite{LLE}. 

\subsubsection*{Notation}

We construct the following matrix
\begin{eqnarray*}
A^j_i&=&\left\{
\begin{array}{cc}
\PM_{i}^{j}&1\le j\le N_p\\
\MM_{i}^{(j-N_p)}&N_p<j\le 2N_p
\end{array}
\right.
\end{eqnarray*}
or in expanded notation
\begin{eqnarray*}
A^j_i&=&\left[
\begin{array}{cccccc}
\PM_{1}^{1} & \cdots  & \PM_{1}^{N_{p}} &
\MM_{1}^{1} & \cdots  & \MM_{1}^{N_{p}}\\
\vdots  & \PM_{i}^{j} & \vdots &
\vdots  & \MM_{i}^{j} & \vdots \\
\PM_{N_{e}}^{1} & \cdots  & \PM_{N_{p}}^{N_{e}}&
\MM_{N_{e}}^{1} & \cdots  & \MM_{N_{p}}^{N_{e}}\\
\end{array}
\right] 
\end{eqnarray*}
which contains the raw, {\em background subtracted} and {\em normalized}
data. By background, we mean fluorescence background, which we identify by
fitting a Gaussian distribution to the subset of all $(PM, MM)$ pairs
satisfying the criterion that $|PM-MM|<\epsilon$, with $\epsilon=50$.
This provides us with the mean and SD in the background fluorescence.
\footnote{for details, see\\
http://xxx.lanl.gov/abs/physics/0102010}. For a fair comparison of compression effects
in various methods, we used the global normalization factors from the MAS 5.0 software
in all cases, however, the technique remains applicable with other normalization
schemes.

\(N_{p}\) is the number of probe pairs and \(N_{e}\) is the
number of experiments.  We introduce a set of weights \( w_{i} \) such
that
\[ \sum _{i=1}^{N_e}\, w_{i}=1 \]
and define the column means (or center of mass)
\[ m^{j}=\sum _{i=1}^{N_e}\, w_{i}\, \logA \]
Note, we are computing the mean of the logs of the components of
$A_i^j$.

\subsubsection*{Local principal component analysis}

\begin{figure}[ht]
\includegraphics[angle=270, width=1\columnwidth]{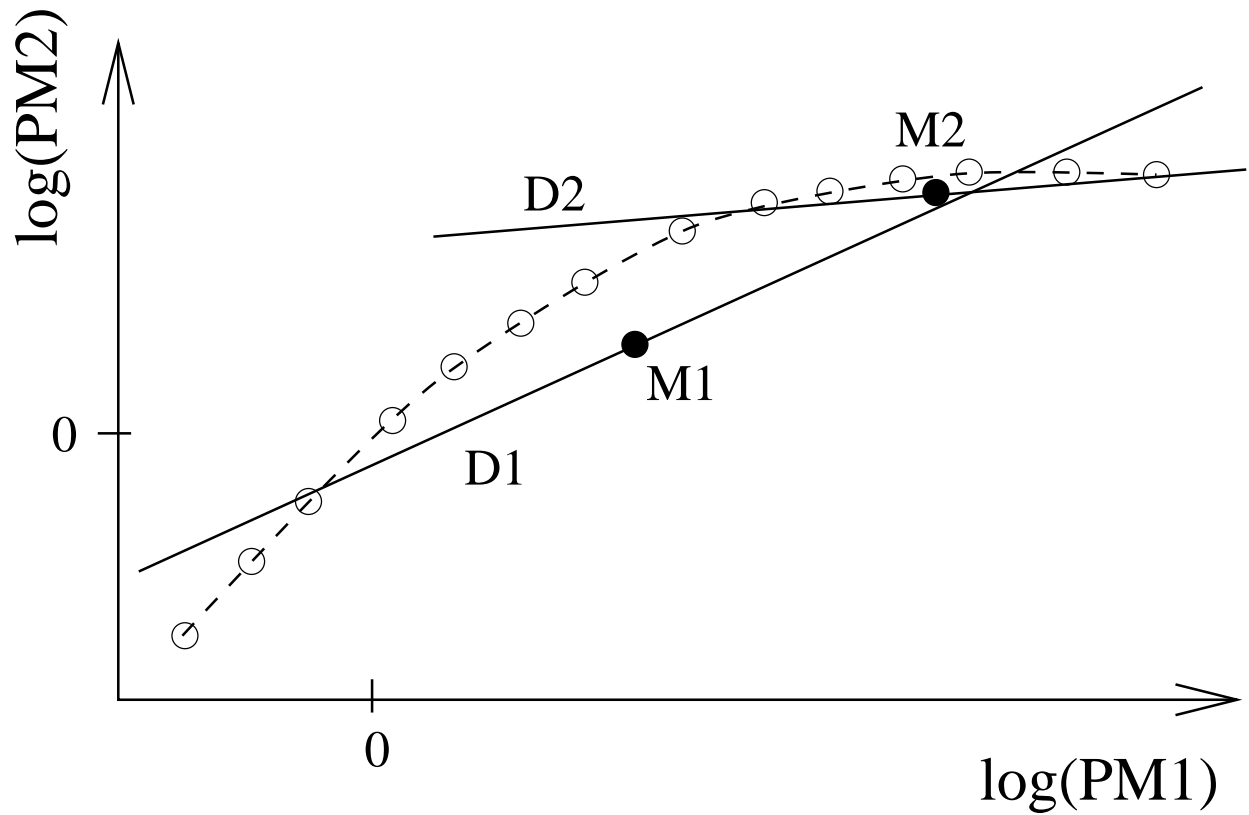}
\caption{\small Typical compressive situation: a 2 dimensional cartoon. The
open dots represent fictitious measurement of two probes (PM1, PM2) at
increasing concentrations (from left to right). Probe PM2 saturates
earlier than PM1. \( M1=(m^{1},\, m^{2}) \) represents the mean with
uniform weights \( \{w_{i}\} \) and M2 a mean obtained with weights
that are larger for high concentrations. D1 and D2 show the
corresponding principal components (direction of largest variance). It
is clear that projecting the points onto D1 has the effect of a
compression due to the curvature. On the other hand, this compression
is largely reduced at high-intensities by projecting onto D2.}
\label{fig:manifold}
\end{figure}

Local embeddings are adequate in situations where compression is important
because non-linearities (resulting from chemical saturation) affect the
one-dimensional manifold \( \{\PM^{j}(c),\, \MM^{j}(c)\} \) (the
concentration \( c \) is the one-dimensional 'curve' parameter) by
giving it a non-zero local curvature. The results section, in particular
Figure~\ref{fig:twoarray} contains ample evidence that these non-linearities
are significant (cf. also Figure 2 in~\cite{chudin}).
Our method is a multidimensional generalization of the schematic depicted
in Figure~\ref{fig:manifold}, 
%
%
%
%
which shows the typical situation of two
probes in which one of the probes (\PM\/2) saturates at concentrations
lower than the other. If both probes were perfectly linear, the
curve would be a straight line with slope 1.
In the multidimensional case, the directions of largest variation
(analogous to D1 or D2 in Figure~\ref{fig:manifold}) are computed from
the principal components of the matrix
\[
\sqrt{{w_{i}}}\, \left( \logA-m^{j}\right) =\sum _{k=1}^{N_p}U_{ik}\,
D_{k}\, V^{j}_{k}
\]
which can easily be done via singular value decomposition (SVD).
In order to reconstruct the concentrations, one
needs first to consider the unspecified sign of the vector \( V^{j}_{1} \)
(when returned by the SVD routine), which has to be chosen such that
the total amount of signal comes out positive. This is easily achieved by
adjusting the sign of \( V^{j}_{1} \) such that
\( \sum _{i,j}\,\logA\, V^{j}_{1}>0 \)~.
The logarithm of the concentration \(s_{i}\), is
then computed by projecting the original matrix onto the first
principal component \( V^{j}_{1} \), corrected by a factor $v_{max}$. This factor
accounts for the fact that the vector \(V^{j}_{1}\) is
\(L_{2}\)-normalized (\(\sum _{j}\,(V^{j}_{1})^{2}=1\)) by definition in the SVD. The signal
then reads:
\[
s_{i}=\vmax\, \sum _{j=1}^{N_p}\logA\, V^{j}_{1}
\]
where
\( \vmax=\max_{j}\left|V^{j}_{1}\right| \). 
In addition, the above procedure automatically yields a signal-to-noise ($S/N$) measure for
the entire probeset
\[
\frac{S}{N}=\frac{D_1}{\sqrt{\sum^{N_p}_{j=2}\,D^2_j}},
\] 
where \(\{D_{k}\}\) are the singular values.
Large $S/N$ values imply that the probeset measurements in the $N_e$ experiments had
a well-defined direction of variation, and can for instance be used as a filter for
identifying genes that exhibit significant changes across the experiments tested.
In the results shown in Figure~\ref{fig:svd}, we used the following weights
\[
w_{i}=\frac{1}{W}\left\{
\begin{array}{cl}
1,&\hat{s}_{i}>\Gamma, \\
1/\!\left({1+b\, (i-\isplit)^{2}}\right), &\hat{s}_{i}\leq \Gamma 
\end{array}
\right.
\]
\noindent 
where \(W=\sum _{i}\,w_{i}\), \(\hat{s}_{i}\) is the signal obtained
with uniform weights, \(\isplit=20\) (out of 28 experiments), \(b=2\),
and \(\Gamma=\hat{S}_{\isplit+1}\) with \(\hat{S}_{i}\) being the
ascendantly sorted $\{\hat{s}_{i}\}$. In other words, lower
concentration points are suppressed according to their rank
(computed with uniform weights) using a slowly decaying Cauchy
weight function. There are of course other weight functions
that could serve the same purpose.

Note that the fitting procedure used in the Li-Wong~\cite{LIWONG} method
is identical to an SVD decomposition, however, with different input
data than was use here. The three main difference between our method
and the Li-Wong technique are: (i) in the analysis here, we used log
transformed \PM\  and \MM\  intensities, rather than the bare \PM-\MM\ 
values; (ii) we introduced optional
weights, which can account for non-linearities of the probe response
in the high concentration regime; (iii) we subtract the column mean
before we compute the principal components, which is crucial for
capturing the local directions of variation.  Indeed, as can be seen
in Figure~\ref{fig:manifold}, the principal component would be
dominated by the mean itself without subtraction.

\subsection*{Results}

\subsubsection*{Data sets}

The yeast Latin square (LS) experiment is a calibration data set
produced by Affymetrix, that uses a non-commercial yeast chip. 14
groups of 8 different genes, all with different probe sets, are spiked
onto 14 chips at concentrations, in pM, corresponding to all cyclic
permutations of the sequence (0, 0.25, 0.5, 1, 2, $\ldots$,
1024). Hence, each group is probed at 13 different finite
concentrations, logarithmically spaced over a range of magnitudes from
1 to 4096 (in Figures~\ref{fig:twoarray} and~\ref{fig:svd}, we refer to
these concentrations as ($1=0.25\,\rm pM$, $2=0.5\,\rm pM$ $\ldots$
$13=1024\,\rm pM$), and each group is completely absent in one array.
Besides the spiked-in target cRNA's, a human RNA background was added
to mimic cross-hybridization effects that would occur in a real
experiment. In addition, each experiment was hybridized twice leading
to 2 groups of 14 arrays called R1 and R2.

The reason why this dataset is attractive as compared to the similar human and e. coli datat available at www.netaffx.com are several.
First, the e. coli data exhibits severe optical saturation, which interfers with the chemical saturation issue we are trying to address here.
The yeast dataset, on the other hand, has virtually no optically saturated cells, as can be inferred the SD in the pixel intensities reported in the raw data (.CEL) files.
In total, fewer than $0.1\%$ of the probes have $SD=0$
($SD=0$ characterizes optical saturated cells).
Further, optical saturation is no longer an issue in GeneChips with
the current scanner settings.
More important, the present dataset permits far better statistics,
as the number of spiked-in genes is 112 as compared to 14 for the human chip. In the latter dataset, there is only one transcript per concentration group as compared to 8 in the yeast case.
In fact, we verified that the compression effects discussed below are virtually identical in the human case (not shown).

%
%
%
%
%

\subsubsection*{Summary of two-array methods}

The figures in the results section show the log-ratios as function of
concentration in the form of boxplots.  In these plots, the central
rectangle indicates the median, 1st and 3rd quartiles (\(Q1\) and
\(Q3\)). The ``whiskers'' extend on either side up to the last point
that is not considered an outlier. Outliers are explicitly drawn and
are defined as laying outside the interval \([Q1-1.5*IQ,\,
Q3+1.5*IQ]\), where \(IQ=Q3-Q1\) is the interquartile distance. For
each method, we show three plots, the top two measure the false
negative rate for ratios of 2 and 4 fold respectively, and the last
one shows the false positive rate. For the top two plots, all
combinations (within R1 and R2 separately) of arrays leading to ratios
of 2 and 4 were considered, and plotted as function of their baseline
intensity (the lesser of the two concentrations). For the third, each
gene was compared between the groups R1 and R2, at identical
concentrations. Of the $8*14=112$ transcripts, 8 were left out of the
analysis because they did not lead to a signal that was tracking the
concentrations at all (presumably due to bad probes or transcript
quality).

\begin{figure}[ht]
\centering
\includegraphics[width=1\columnwidth, angle=0]{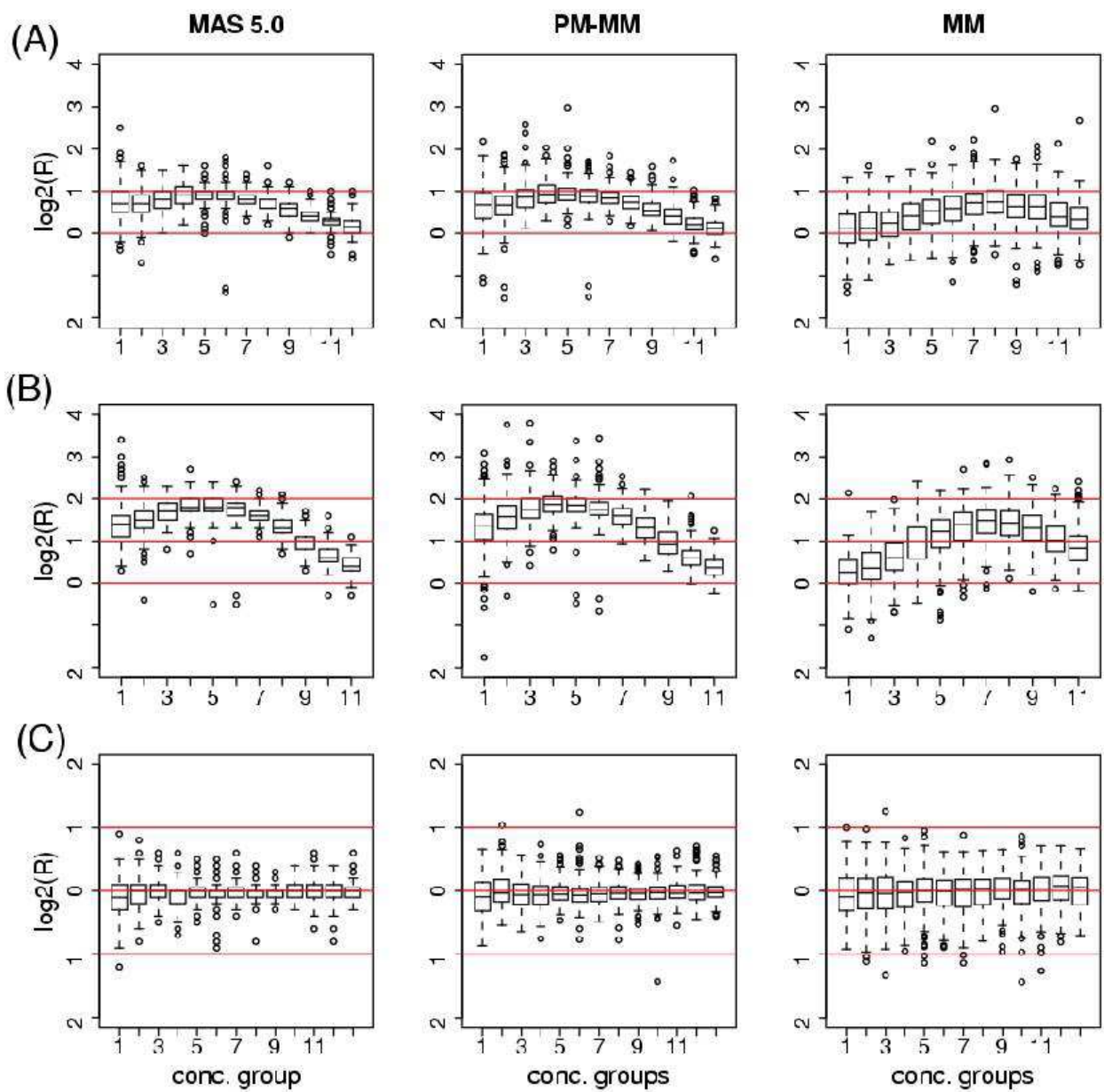}
\caption{\small Comparison of 'two array' methods: MAS 5.0 (first column),
$\PM-\MM$ (second) and \MM only (third) of~\cite{US3}.  Boxplots show
the log base 2 ratio distributions for each baseline concentration group
(cf. text). Row A:
Fold change of 2, B: Fold change 4, C: Negative controls (false
positives).  The central rectangle indicates the median, 1st and 3rd
quartiles (\(Q1\) and \(Q3\)). The ``whiskers'' extend on either side
up to the last point which is not considered to be an
outlier. Outliers are explicitly drawn and are defined as laying
outside the interval \( [Q1-1.5*IQ,\,Q3+1.5*IQ] \), where \( IQ=Q3-Q1
\) is the interquartile distance. Notice the two first rows are
qualitatively similar, with the MAS 5.0 being marginally cleaner. Both
methods show a strong high concentration compression, but have
excellent reproducibility (cf. text). The third column illustrates
that MM probes contain valuable signal, often leading to more accurate
ratios at high concentrations.}
\label{fig:twoarray}
\end{figure}

\begin{figure}[ht]
\centering
\includegraphics[width=1\columnwidth, angle=0]{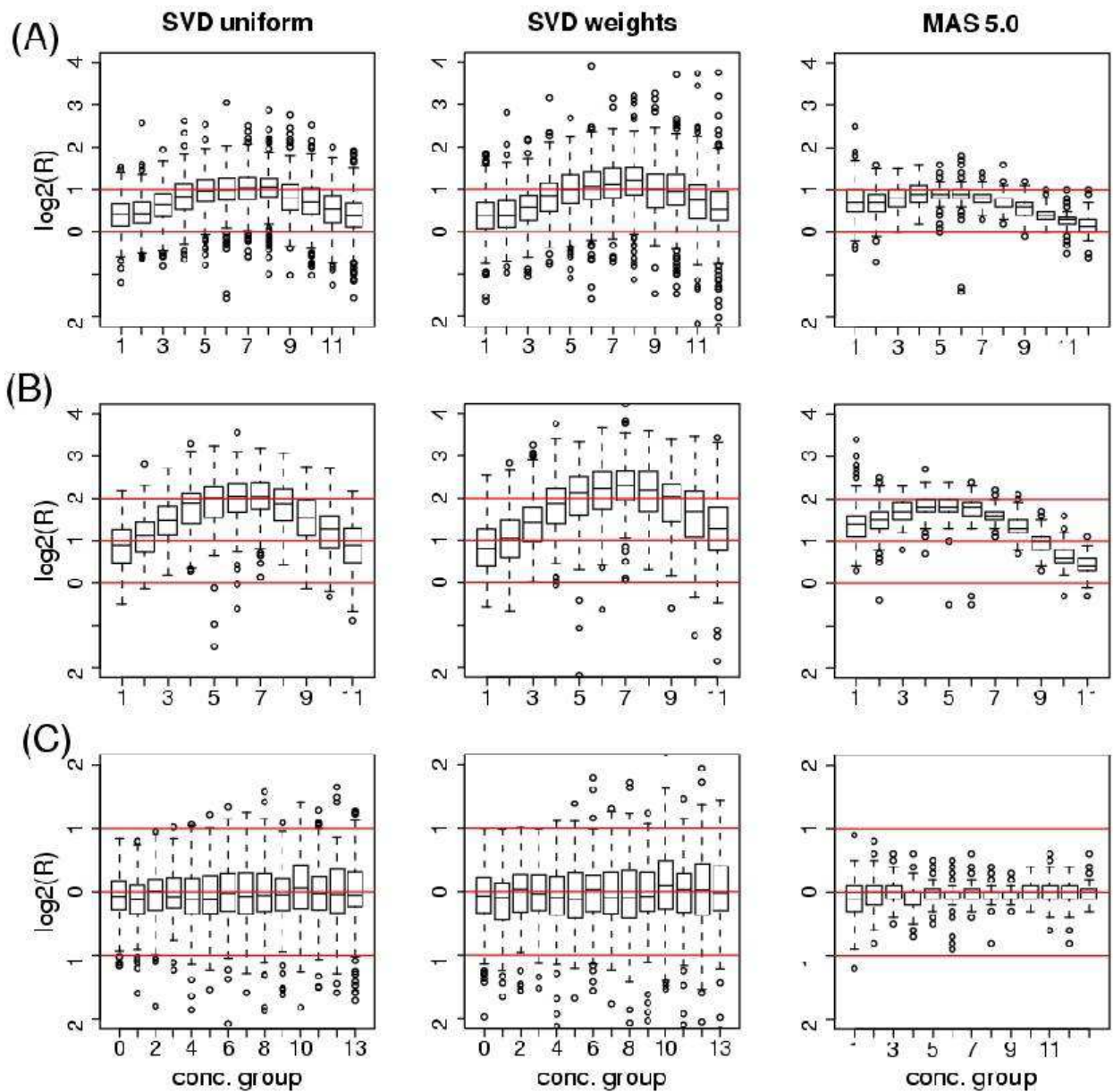}
\caption{\small Rows are as in Figure~\protect\ref{fig:twoarray}, columns show
the new method with uniform (first column), Cauchy weights introduced
in the text (second), and the reference MAS5.0 (third).}
\label{fig:svd}
\end{figure}

In Figure~\ref{fig:twoarray}, we summarize the results obtained by
the Microarray Analysis Suite 5.0 (MAS 5.0) software and the ``2
chips'' method discussed in~\cite{US3}. The later method computes for
each gene probed in two arrays a ratio score \(R\) such that
\[ \log (R)=\sum _{j}^{robust}\log (R^{j}) \]
is a robust geometric mean (a least trimmed squares estimator) of the
probe ratios \(R^{j}\). Figure~\ref{fig:twoarray} shows the cases
where
\[
R^{j}=\frac{\PM^{j}_{1}-\MM_{1}^{j}}{\PM^{j}_{2}-\MM^{j}_{2}} 
\] 
and
\[
R^{j}=\frac{\MM_{1}^{j}}{\MM^{j}_{2}} 
\]
In both cases, only probes with numerator and denominator above
background are retained. The first case ($\PM-\MM$) is in essence
similar to the MAS 5.0 program, differences are in the choice of the
Tuckey bi-weight mean as the robust estimator, and in the treatment of negative differences. For our
purpose here, we like to think of the Affymetrix method as
two-array, $\PM-\MM$ based method. In all the results presented below,
the arrays were scaled according to the MAS 5.0 default settings.

The main features of Figure~\ref{fig:twoarray} are: there is an
optimal range of baseline concentrations ($\approx 1-16$ pM) in which
the ratio values from both $\PM-\MM$ methods (the two first columns)
are fairly accurate, for both ratios of 2 and 4. For both lower and
higher concentrations, there is a noticeable compression effect, which
is most dramatic at the high end.  At the highest baseline
concentration (512 pM for the ratios of 2 and 256 pM for ratios of 4),
changes of 2 are basically not detected and real changes of 4 are
compressed on average to values around 1.25. The analysis of the false
positive rate (last row) shows that both methods yield very tight
reproducibility: the log2 ratio distributions are well centered around
0 and the interquartile distances are roughly intensity independent
and smaller than $~0.2$, meaning that 50\% of the measurements fall in
the ratio interval $[0.93, 1.07]$.  To be fair, we should point out
that as a ($\PM-\MM$\/) method, the MAS 5.0 algorithm is on average a
bit cleaner, having slightly fewer outliers.  However, we like to
emphasize that the qualitative behavior in the two ($\PM-\MM$\/) methods
is unchanged, especially as far as the high-intensity compression is
concerned. Further, similar behavior is also found using the
($\PM-\MM$\/) Li-Wong method (data not shown).  The above observations
are consistent with what was reported in~\cite{chudin}, confirming that
these effects are independent of the chip series.

The third column in Figure~\ref{fig:twoarray} illustrates our
contention that the \MM\ are in essence a set of lower affinity
probes.  We notice that using only the \MM\ measurements in the
two-array method changes the picture qualitatively.  Whereas the low
concentration regime is far worse than in the ($\PM-\MM$\/) methods, the
behavior toward the high end has changed and the drop off occurs now
at higher concentrations: approximately 256 pM for the ratios of 2 and
128 pM for ratios of 4. On the other hand, even in the optimal range,
the magnitude of the medians are always a bit lower than the real
ratios, and the false positive rate also suffers. To summarize, this
result suggests that if one is interested in accuracy at high
concentrations, then the \MM-only methods offers the best two-array
alternative. We have tried other variations: \PM\ only, or the double
size set consisting in the merged \PM\ and \MM's, both being worse at
high concentrations than the \MM\ only method.

\subsubsection*{Multi-array methods}

The data analyzed using our new method is shown in
Figure~\ref{fig:svd}. 
%
%
%
%
It is clear that both are capable of reducing
the high intensity compression, as compared to existing methods. The
second column explicitly shows the higher accuracy of the local method.
It should be noted, however, that the precision is
significantly lower than with MAS 5.0, which is the trade-off to pay for
higher signal detection. As compared to the ``two chip'' \MM\ method,
which was previously the least compressive in this regime,
the medians are systematically more accurate.
Also, the method does not perform well at low-concentrations
which is expected since it was not designed for that range. 

\subsubsection*{Significance scores}
\begin{figure}[ht]
\centering 
\includegraphics[width=1\columnwidth, angle=0]{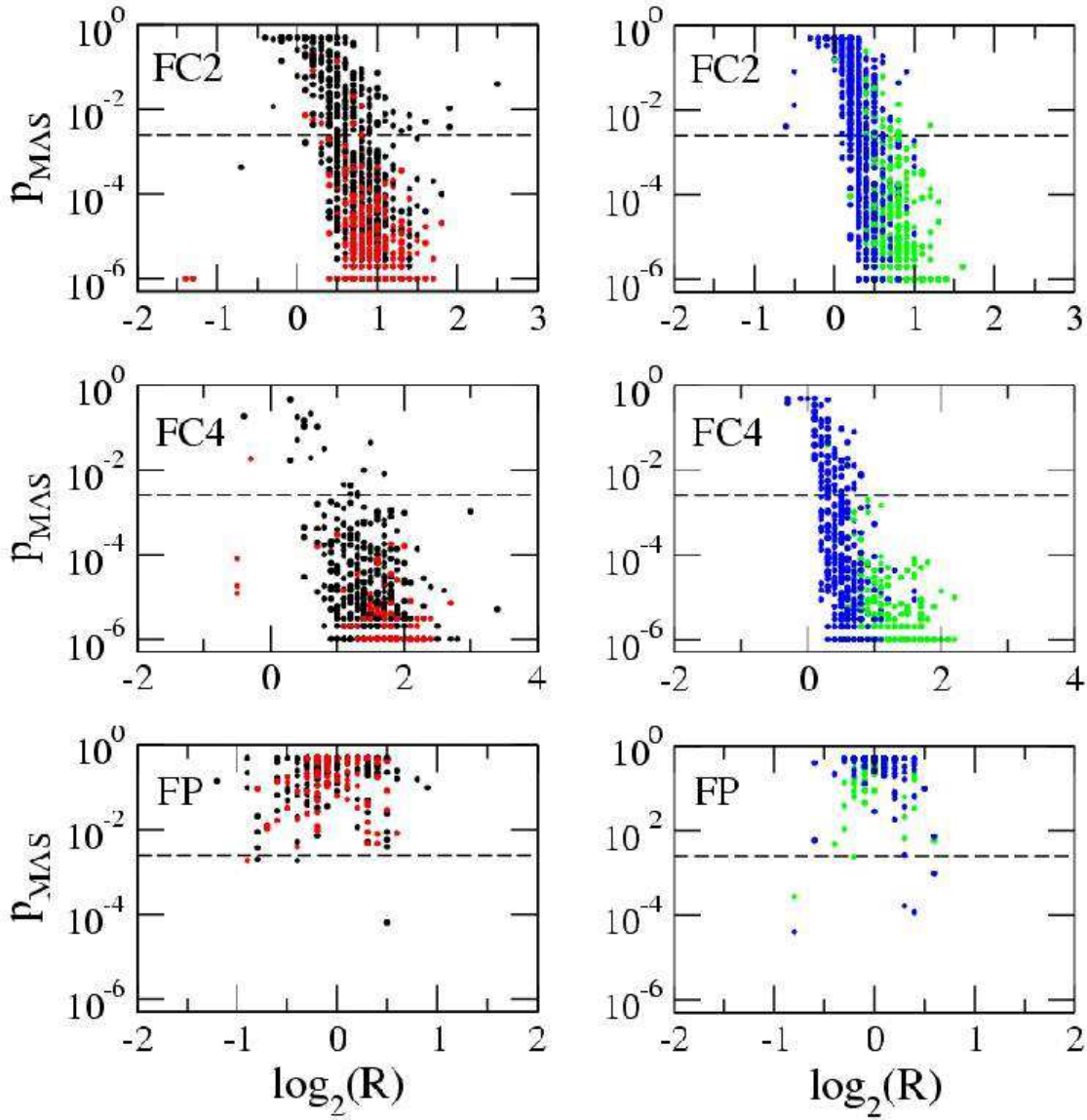}
\caption{\small P-values vs. log-ratios in for MAS 5.0. The plotted
$p_{MAS}$ is the transformed MAS 5.0 p-value (cf. text). The dotted
line indicates the default $\gamma_1=0.0025$, below which MAS 5.0
scores are considered increased (I) or decreased (D) (for the
transformed p-value).  Colors are used to group baseline intensities
of 0.25--1 pM (black), 2--8 pM (red), 16--64 pM (green), 128--1024 (blue).}
\label{fig:pvals}
\end{figure}

Although ratio score may suffer severe compression, there remains the
possibility that they would be attributed a significant increase or
decrease call. 
%
%
%
%
Figure~\ref{fig:pvals} displays the relation between
the MAS 5.0 log-ratios and their associated p-values. MAS 5.0 change
`p-values' $p_M$ are symmetric about 0.5 and designed such that the
ratio score is called increased when $p_M<\gamma_1$ and decreased (D)
when $p_M>1-\gamma_1$, with a default $\gamma_1=0.0025$. This
definition is not well suited for plotting purposes, we therefore work
with $p_{\it MAS}=p_M$ when $p_M<0.5$, and $p_{\it MAS}=1-p_M$
otherwise. This way, both I and D genes have $p_{\it MAS}<\gamma_1$,
the direction being given by the sign of the log-ratio.  The results
show that there remarkably few false positive calls: only 4 out of 624
for concentrations $c\leq 8\,\rm pM$, and 6 of 728 when $c\geq 16\,\rm
pM$. Fold changes of 4 are also well detected despite the compression
at high intensities: there are 21 false negatives (and 3 false
positives having ratios with the wrong sign) out of 1248 for $c\leq
8\,\rm pM$, and 84 of 1040 false negatives for $c\geq 16\,\rm pM$. The
situation deteriorates for fold changes of 2, with 124 false negatives
(and 3 false positives) out of 1248 for $c\leq 8\,\rm pM$, and 425 of
1248 false negatives for $c\geq 16\,\rm pM$.

\subsection*{Discussion}

We have shown that high-concentration bias is a serious issue
in GeneChips, which is probably related to chemical saturation in the adsorption process of the target to the probes.
Exploiting the broad range of affinities of different probes (\PM\ and \MM\ included) offers an approach toward improvements.
However, the gain in accuracy comes with an expected decrease in precision, since effectively, the weight of a measurement is transferred to a {\em smaller} set of probes.
Hence, the reduction in noise levels resulting from averaging over
probes is diminished.

Our method should serve as a useful complement to those who use
microarrays primarily as a gene discovery tool, and are interested
in maximal signal detection.
In fact, we often hear that severe constraints like pharmaceutical
treatments or gene knockouts appear to have no detectable transcriptional effects.
While there is the possibility that transcription regulatory
networks can compensate for such changes, or that some
effects would be mostly post-transcriptional, real transcriptional
changes may also be masked by compressive effects like those discussed
above.

\subsection*{Conclusion}

We have summarized the performance of existing methods for
analyzing Affymetrix GeneChip data, using the yeast calibration
dataset from Affymetrix. The results show
unambiguously the compressive tendency of GeneChip measurements in the
high-intensity range, namely that fold changes as large as 4 in
expression levels can be reduced to fold changes barely larger than 1
(Figure~\ref{fig:twoarray}). Interestingly, we showed that among the standard techniques, the one using only the \MM\ signals offers the highest accuracy at high concentrations.
Additionally, we have how it is possible to achieve higher accuracy at high concentrations by exploiting the probeset's wide affinity range. One should realize, however, that saturation problems of the sort
encountered present a hard challenge in signal processing, and it is therefore expected that higher accuracy is obtained at the expense of
reduced precision.

Our observations raise the following design issue for oligonucleotide arrays. Since it will likely be difficult to manufacture oligonucleotide probes with linear responses over 4 or more decades in concentration, an option would be to optimize the design of probesets such that each of its probe would be optimally linear in smaller ranges
(say 2 decades at most) centered around graded concentrations. In this way, the weights of a measurement could be transferred to an appropriate subset of probes known to be optimal in a given range.
Hence, one would use a different set of probes for high or low concentration values to increase the overall dynamic range of a probeset.

\section*{Acknowledgments}

The authors are thankful to Affymetrix for having provided
the useful calibration data. F.~N. is a Bristol-Meyers Squibb Fellow and
acknowledges support from the Swiss National Science Foundation.
This work was also supported by NIH/NINDS Grant NS39662 (N.~D.~S)
and the Meyer Foundation (M.~M.)

\bibliographystyle{bioinformatics}
\bibliography{microarray}

\onecolumn

\subsection*{Figures}

\vfill

\pagebreak
\vfill

\pagebreak
\vfill
\pagebreak
\vfill

\pagebreak
\vfill



\end{document}